\documentclass[iicol]{sn-jnl}

\usepackage{epsfig,amssymb,amsfonts,amsmath,mathtools,bm,color,graphicx,orcidlink,bm,braket,multirow,dsfont,mathtools,xcolor,slashed}

\RequirePackage[numbers,sort&compress]{natbib}
\renewcommand{\vec}[1]{\mbox{\boldmath$#1$\unboldmath}}

\newcommand{\be}{\begin{equation}}
\newcommand{\ee}{\end{equation}}
\newcommand{\bea}{\begin{eqnarray}}
\newcommand{\eea}{\end{eqnarray}}
\newcommand{\beas}{\begin{eqnarray*}}
\newcommand{\eeas}{\end{eqnarray*}}

\def\vec#1{{\bm #1}}

\graphicspath{{Figs.dir/}}

\begin{document}

\title{Three regimes/phases of QCD at high T, their symmetries and $N_c$ scaling}

\author{L.~Ya.~Glozman}

\affil{\orgdiv{Institute of Physics}, \orgname{University of Graz}, \orgaddress{ \postcode{8010}, \city{Graz}, \country{Austria}}}

%\PACS{{12.38.Aw}{} \and {12.39.Ki}{} \and {11.30.Rd}{}}

\date{}

\abstract{We review recent developments on the QCD phase
diagram at small chemical potentials and increasing temperature.
There are three regimes/phases in QCD which differ by symmetries,
degrees of freedom and $N_c$ scaling: the hadron gas below the chiral restoration temperature $T_{ch}$, the stringy
fluid  between $T_{ch}$ and the deconfinement temperature $T_d$  and the quark-gluon plasma above $T_d$. 
} %end of abstract
\maketitle

\section{Introduction}

In this talk we overview recent   progress
in  understanding of the QCD phase diagram at small
chemical potentials and increasing temperature. A more
detailed exposition can be found in two  complementary
reviews \cite{G2,G3}.	

\section{Chiral spin symmetry in QCD and  its flavor extensions. Observation in  the vacuum.}

The QCD equation of motion prescribes  the interaction of the color charge
of a particle with the chromoelectric field to be of the same form as the interaction of the
electric charge with the electric field in electrodynamics:
\begin{equation}
 \vec F = Q^a \vec E^a, ~~ Q^a = g\int d^3x ~ 
\psi^\dagger(x) T^a \psi(x),
\end{equation}
with $T^a,~~a=1,...,8$ being the $SU(3)$ color generators and $g$ is the strong
coupling constant.
The color charge of quarks is invariant under the $SU(N_F)_R \times SU(N_F)_L \times U(1)_A$ chiral symmetry transformations of the Dirac Lagrangian. However it is also
invariant under the chiral spin 
$SU(2)_{CS}$ transformation that mixes the right- and left-handed Weyl spinors
\cite{G1,GP}.
The Dirac Lagrangian 
as well as the interaction of the spatial current of quarks with the chromomagnetic field
are not invariant under the $SU(2)_{CS}$.
 The product of the chiral group with the 
$SU(2)_{CS}$ group can be embedded into a $SU(2N_F)$ group, which is
also a symmetry of the color charge and of the chromoelectric interaction
with light quarks.

The chiral spin and $SU(2N_F)$ symmetries  are broken by the quark kinetic term, by the magnetic interaction,
by the $U(1)_A$ anomaly and by the quark condensate. They can be seen
as approximate symmetries in observables if, and only if the physics 
is dominated by the chromoelectric interaction. In QCD
the chromoelectric interaction is considered to be crucial for confinement of quarks,
because it is well established with the static quarks that confinement
is related to the linear interquark potential, which is connected to the area law of the Wilson loop.  Then it is natural to consider the $SU(2)_{CS}$ and $SU(2N_F)$
symmetries as symmetries of the confining interaction with ultrarelativistic
light quarks. Of course, in order to discuss
the electric and magnetic components of the gauge field one must fix a reference frame. The invariant hadron mass is the rest frame energy.
Consequently we usually choose the hadron rest frame to discuss
physics of the hadron mass generation. In a medium at high temperatures
the Lorentz invariance is broken and the preferred frame is the medium rest
frame.
The full symmetry of confinement in QCD with light quarks
is $SU(2N_F) \times SU(2N_F)$ \cite{G2,G3}.

\begin{figure}
\centering 
\includegraphics[width=0.8\linewidth]{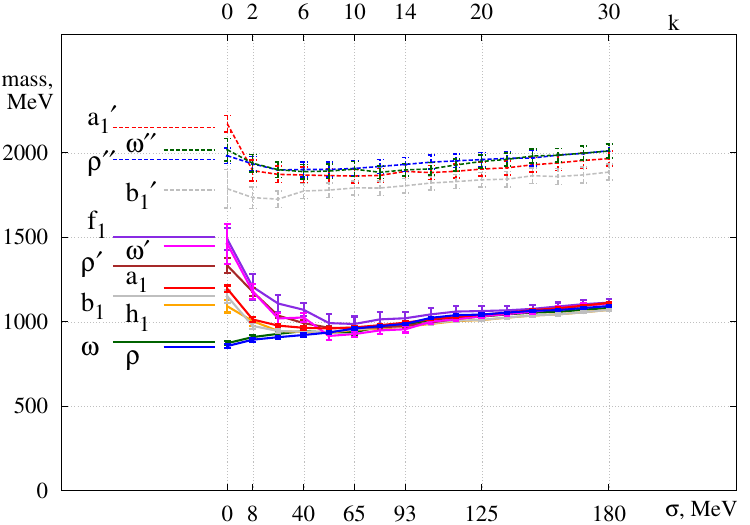}
\caption{$J=1$ isovector and isoscalar meson masses  as a function of the
truncation number $k$ where $k$ represents the amount of removed
lowest modes of the Dirac operator. $\sigma$ shows the energy gap in the Dirac spectrum. From Ref. \cite{D2}.} 
\label{den}
\end{figure}

The $SU(2)_{CS}$, $SU(4)$ and $SU(4) \times SU(4)$ symmetries 
were  observed on the lattice upon the artificial restoration
of chiral symmetry in hadrons (through  truncation of the near-zero modes of the Dirac
operator) in Refs. \cite{D1,D2,DGP1,DGP2}.
E.g. in Fig \ref{den} we see a clear approximate degeneracy of $J=1$ mesons
within the $SU(4) \times SU(4)$ multiplet in the chirally symmetric world.
We conclude:

$\bullet$ While the quark condensate of the vacuum does contribute
 to some extent to the hadron mass, it would be incorrect to say
 that the hadron mass comes from the condensate, i.e. is a consequence of spontaneous breaking of chiral symmetry. The role of the chiral symmetry
 breaking is to lift the $SU(4) \times SU(4)$ degeneracy of the confining electric
 interaction.

$\bullet$ Confinement and spontaneous breaking of chiral symmetry
are not directly related phenomena: the elimination of the chiral symmetry breaking in a medium at high temperatures does not require deconfinement. 

\section{Observation of the $SU(4)$ above $T_{ch}$ and its implications}
 
 Above $T_{ch} \sim 155$ MeV the $SU(2)_L \times SU(2)_R$ chiral symmetry is restored. The $U(1)_A$ symmetry is still broken by the anomaly; the latter breaking is however small.
In the right panel of  Fig. 2 we show temporal correlators of isovector
$J=0,1$ mesonic excitations at $T=220$ MeV in $N_F=2$ QCD calculated with the
chirally symmetric Dirac operator at physical quark masses \cite{R3}.
 \begin{figure}
\centering 
  \includegraphics[width=0.49\linewidth]{{{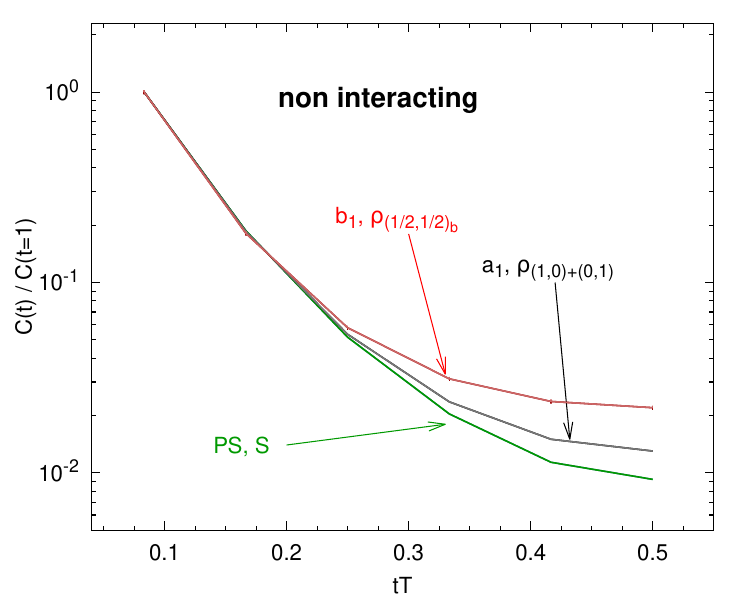}}}
  \includegraphics[width=0.49\linewidth]{{{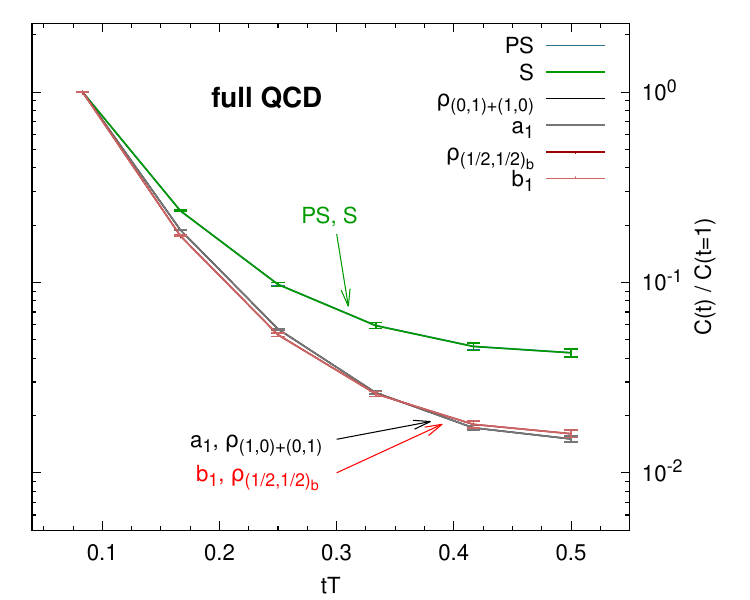}}}
    \label{tcorrs}
\caption{ Temporal correlation functions for $12 \times 48^3$
lattices. The l.h.s. shows correlators calculated with free
noninteracting quarks with manifest $U(1)_A$  and $SU(2)_L \times SU(2)_R$
symmetries. The r.h.s. presents full QCD results at a temperature 220 MeV,
which shows multiplets of all  $U(1)_A$, $SU(2)_L \times SU(2)_R$, $SU(2)_{CS}$  and $SU(4)$ groups. From Ref. \cite{R3}.
}
\end{figure}
The (approximate) degeneracy of the $J=0$ isovector scalar (S) and pseudoscalar (P)
correlators reflects the approximately restored $U(1)_A$ symmetry.
 The approximate
degeneracy of the isovector $J=1$ correlators $a_1$, $\rho_{(1,0)+(0,1)}$,
$b_1$, $\rho_{(1/2,1/2)_b}$ indicates emerged $SU(2)_{CS}$ and
$SU(4)$ symmetries.

We can compare the correlators in full QCD with the correlators
calculated on the same lattice without the gluonic field, see left panel of
 Fig.~2. The latter correlators represent a free
 quark gas and would correspond to the quark-gluon plasma at a very
 high temperature. 
 In the free quark gas the  $U(1)_A$ and $SU(2)_R \times SU(2)_L$ chiral symmetries are manifest. A qualitative difference between the pattern on the l.h.s and the pattern on the r.h.s of Fig.~2 is
 appealing. In full QCD at $T=220$ MeV one observes not only expected
$U(1)_A$ and $SU(2)_R \times SU(2)_L$  symmetries, but also approximate
emerged  $SU(2)_{CS}$ and
$SU(4)$ symmetries.

\begin{figure}
\centering 
  \includegraphics[width=0.49\linewidth]{{{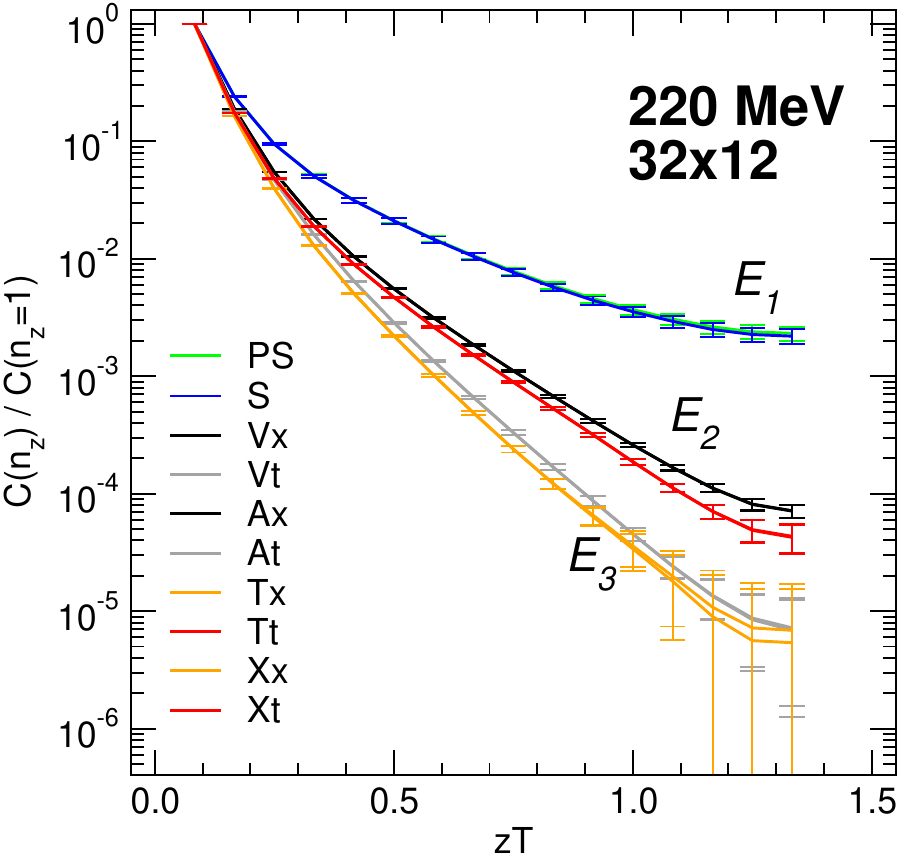}}}
  \includegraphics[width=0.49\linewidth]{{{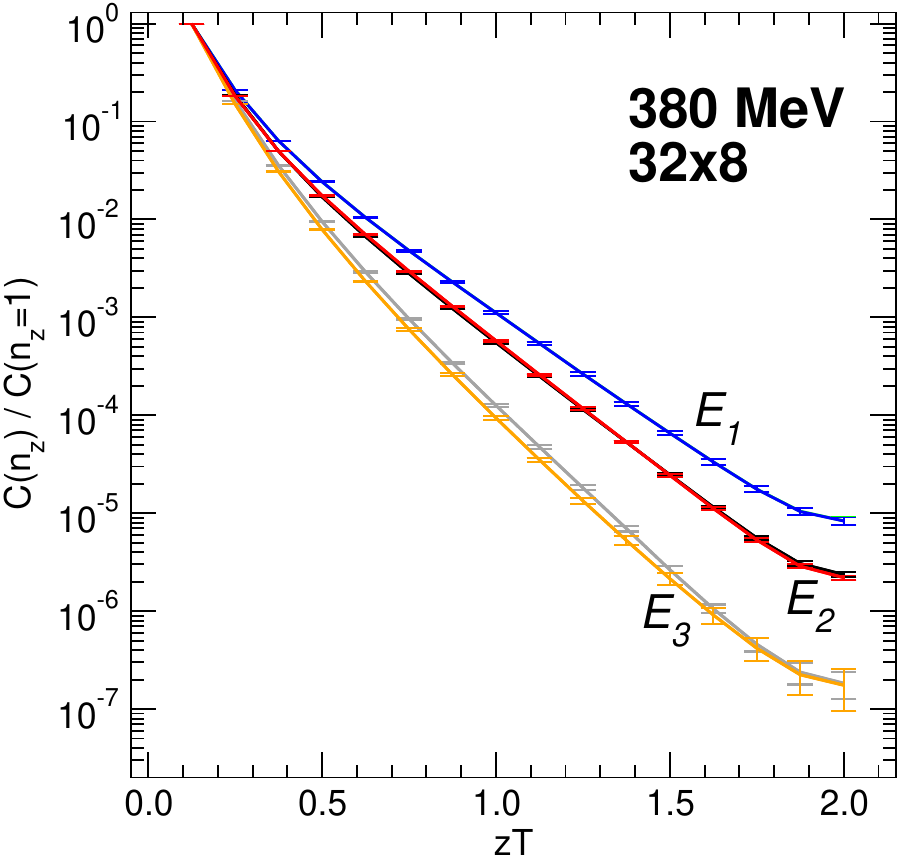}}}
\caption{Spatial
correlation functions of all possible isovector $J=0,1$ bilinears.
  From
Ref. \cite{R2}.}
\label{spatial}
\end{figure}

Now we will discuss observed symmetries of the spatial
correlators at different temperatures \cite{R1,R2}. 
A complete set of all  isovector
 $J=0,1$ correlators is shown in Fig. ~\ref{spatial}.
We see a distinct multiplet structure of the correlators.
The multiplet $E_1$ consists of isovector scalar (S) and pseudoscalar (PS)
correlators and reflects the approximately restored $U(1)_A$ symmetry.  
The multiplet $E_2$ contains four approximately
degenerate correlators obtained with  
 $J=1$ isovector operators. 
All four correlators 
are connected by the $SU(4)$ transformation. 
The multiplet $E_3$ represents the $z$-direction propagation of conserved
charges. The conserved charges do not propagate in time.
 The degeneracy of the normalized
correlators in the $E_3$ multiplet is consistent
with chiral symmetry alone.  
We observe approximate emerged $SU(2)_{CS}$ and $SU(4)$ symmetry up
to temperatures of about $\sim 500$ MeV. At higher temperatures
the distinct multiplet $E_2$ disappears. This would
naturally happen because of the Debye screening of the electric confining interaction. Similar results have been obtained in $N_F=2+1+1$ QCD \cite{Chiu}.
 
We conclude that the QCD thermal partition function above 
the chiral crossover has not only chiral symmetries
but is approximately symmetric with respect to
$SU(2)_{CS}$ chiral spin group and its flavor extension $SU(4)$.
  This implies that the medium is not
a quark gluon plasma which is a system of weakly interacting
(quasi)partons and where only chiral symmetries exist.
The emergent $SU(2)_{CS}$ and $SU(4)$ symmetries 
suggest that the physical propagating degrees of freedom at these temperatures
are chirally symmetric quarks bound into color singlets by the
confining chromoelectric field.
The stringy fluid regime arises above $T_{ch}$ and extends to
roughly $3 T_{ch}$.
Above these temperatures the chiral spin symmetry 
disappears because the confining electric field gets screened
and one observes  a quark-gluon plasma.

The symmetry data suggests that the deconfinement crossover is very smooth.
One might assume that it is a crossover around the temperature $T_d \sim 300$ MeV of the deconfinement phase transition of the pure glue theory \cite{CG1}.
This assumption is supported by the deconfinement temperature obtained from
the center vortices percolation in QCD \cite{Mickley} as well as by a parameter-free
 Hagedorn temperature $T_H \sim 300$ MeV within the string
description \cite{Fujimoto,Mar}. Further evidence for the intermediate regime/phase can be found in Ref. \cite{GPP}, where it is shown that
the perturbative expansion in hot QCD fails below $T \sim 600$ MeV. Ref. \cite{LP} demonstrates that the pion spectral function above $T_{ch}$ exhibits
clear $\pi,\pi'$ peaks. The bottomonium spectrum at $ T < 300$ MeV turns out to
be the same as in vacuum; the states only become broader with temperature \cite{Ding}.

\section{Three regimes/phases of QCD and their $N_c$ scaling}

\begin{figure}
\centering
 \includegraphics[width=0.65 \linewidth]{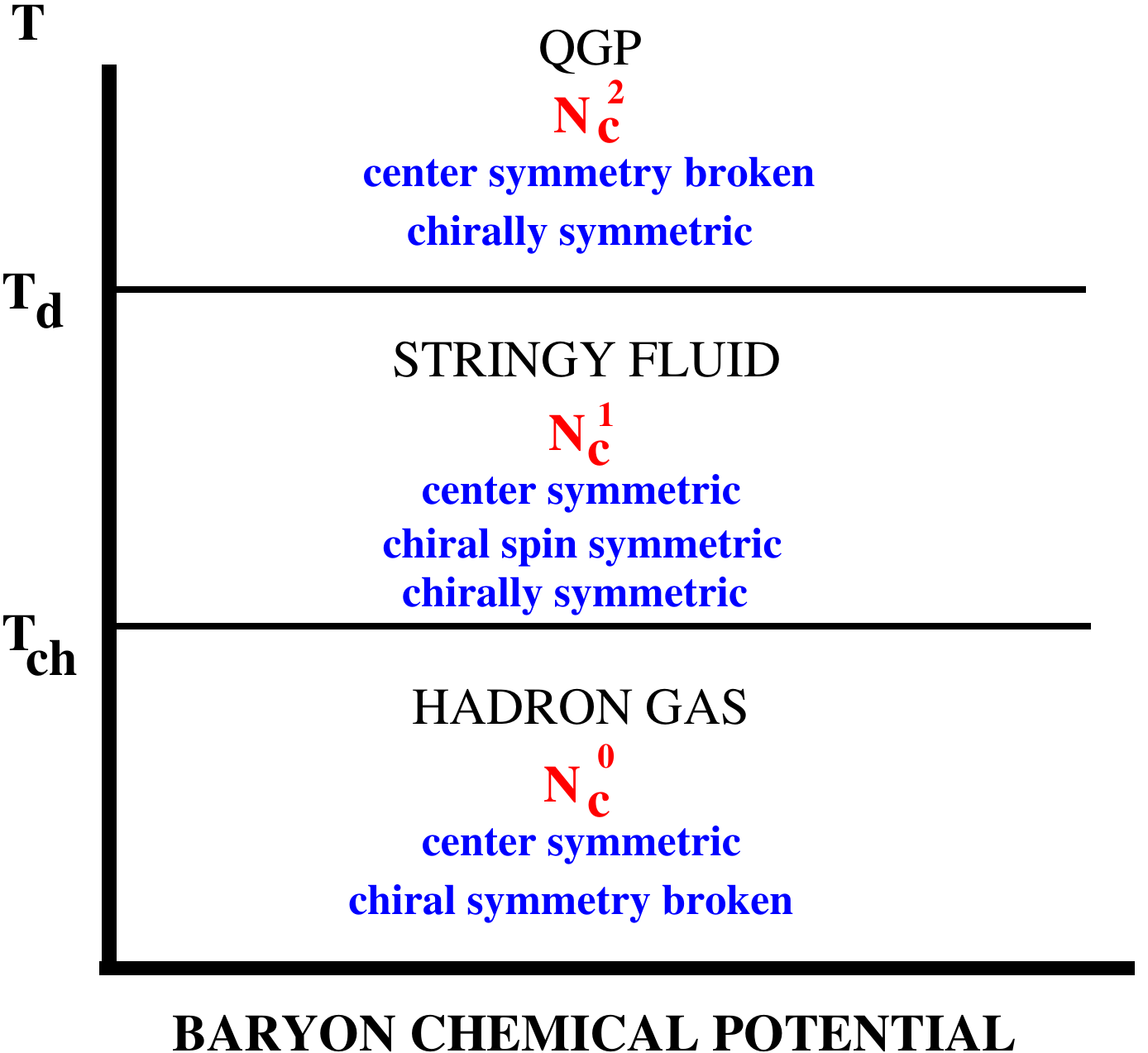}
 \caption{Three regimes/phases of QCD at small chemical potential
with their symmetries and $N_c$ scaling. With a sufficiently large $N_c$ 
the quark loops are suppressed  and the center symmetry is a  relevant symmetry; the order parameter for deconfinement/confinement  is then the Polyakov loop. In the large $N_c$
limit one expects that the stringy fluid regime becomes a distinct phase
separated from the hadron gas and the QGP by phase transitions.}
\label{diag}
\end{figure}

It has been suggested in ref. \cite{CG1} that the hadron gas, the stringy fluid and the quark-gluon plasma are
 characterized by different scaling of the energy density $\epsilon$, pressure $P$
and entropy density $s$ with $N_c$, where $N_c$ is the number of colors in QCD,
see Fig. ~\ref{diag}:
\begin{equation}
    \epsilon_{\rm HG} \sim N_c^0 \; \; , \; \; P_{\rm HG} \sim N_c^0 \; , \; s_{\rm HG} \sim N_c^0 \; , \label{Eq:had scale}
\end{equation}
 
\begin{equation}
    \epsilon_{\rm str} \sim N_c^1 \; \; , \; \; P_{\rm str} \sim N_c^1 \;  ,  \; s_{\rm str} \sim N_c^1 \;  ,
     \label{Eq:IntScal}
\end{equation}

\begin{equation}
    \epsilon_{\rm QGP} \sim N_c^2 \; \; , \; \; P_{\rm QGP} \sim N_c^2 \;  ,  \; s_{\rm QGP} \sim N_c^2 \;  .
     \label{Eq:IntScal}
\end{equation}
The $N_c^0$ scaling of the thermodynamic quantities 
in the hadron gas  as well as the $N_c^2$ scaling within the QGP \footnote{In the deconfined phase there are
$N_c^2-1$ independent gluon quasiparticles and only $N_c$ quark quasiparticles. Consequently  it is the gluon quasiparticles which dominate
the thermodynamic quantities in the equilibrium.}
were understood long ago. 
 The physical reason for the $N_c^1$ scaling
within the confined but chirally symmetric stringy fluid   is  the following \cite{G4}. The  energy of the color-singlet quark-antiquark
excitations scales as $E_k \sim N_c^0$.
At vanishing chemical potentials all expectation values 
of the bilinear color-singlet operators $\bar q \Gamma_k q$ (where $\Gamma_k$ includes $\gamma$- and flavor-matrices) with not vacuum quantum numbers automatically vanish,
$<\bar q \Gamma_k q>=0$.
 However, fluctuations of these quantities
do not vanish. 
So we estimate the energy density in a dense medium of the overlapping color-singlet systems as a product  of the fluctuations   of such color-singlet quark-antiquark pairs,
 \begin{equation}
 n_k =\sqrt{\int d^3 x_1 d^3 x_2 <\bar q(x_1) \Gamma_k q(x_1) \bar q(x_2) \Gamma_k q(x_2)>} ,
 \label{fl}
\end{equation}
 and the energy  $E_k$ of each pair:
\begin{equation} 
\epsilon_{\rm str} = \sum_k n_k E_k  \sim N_c^1,
\label{de}
\end{equation}
because $n_k \sim N_c^1$.

The different $N_c$ scaling of the three regimes has important
consequence for the structure of the phase diagram. When $N_c$
is sufficiently  large one expects that the energy density experiences  jumps at $T_{ch}$ and $T_{d}$ and instead of three regimes with smooth crossovers in the real world $N_c=3$ one obtains three
distinct phases with first order phase transitions between them \cite{CG1}.

The three regimes/phases picture of Fig.~\ref{diag} should be a
subject for  lattice studies. The equation of state 
at $N_c=3$ has been explored \cite{Bor,hot}. One should repeat  measurements
of the thermodynamic quantities at $N_c > 3$
and establish the $N_c$ scalings at different temperatures. 

\section{Fluctuations of conserved charges as evidence for  stringy fluid}
In the hadron gas regime/phase  below $T_{ch}$ 
the hadron structure is not yet resolved and its internal degrees of freedom
are frozen, which is the reason for the $N_c^0$ scaling. 
Once the density of hadrons increases and they start to overlap the internal hadron structure gets relevant.
\begin{figure}[h]
\centering
 \includegraphics[width=0.48 \linewidth]{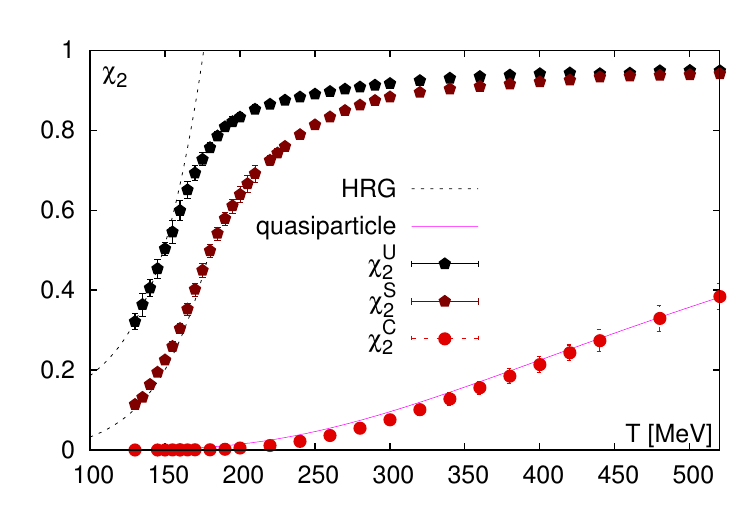}
  \includegraphics[width=0.48 \linewidth]{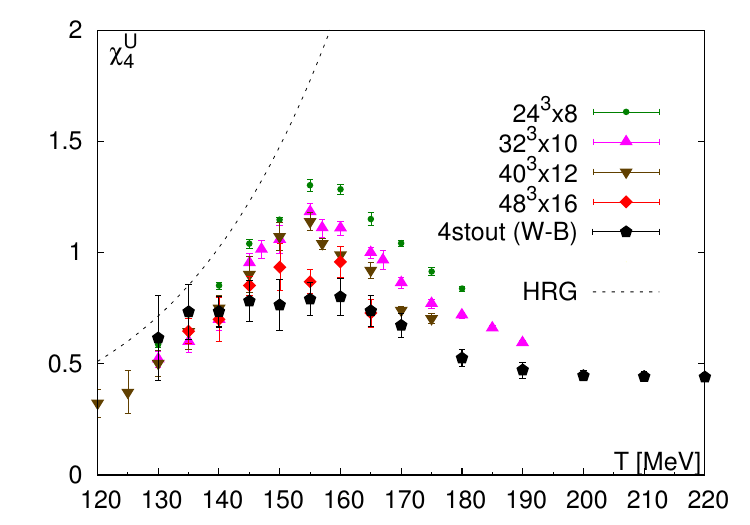} 
  \caption{Left panel: Fluctuations of conserved net $u$ ($\chi_2^U= \chi_{2,0,0}^{u,d,s}$) and strange ($\chi_2^S= \chi_{0,0,2}^{u,d,s}$) quark numbers in 2+1 QCD at physical quark masses. $\chi_2^C$
    is irrelevant to our discussion and can be ignored. 
    Right panel: Cumulant $\chi_4^U= \chi_{4,0,0}^{u,d,s}$  in 2+1 QCD at physical quark masses. From. Ref. \cite{Bel}.  }
    \label{fluc}
\end{figure}
Consider charges associated with the net number of u,d,s quarks:
\begin{equation}    
N_q \equiv \int d^3 x ~n_q(x) \; \;\;  {\rm with}
\; \; \;n_q(x) = \bar q(x) \gamma^0 q(x), \;\; \; q=u,d,s
\label{def}
\end{equation}
The  conserved
flavor charges  $N_q$ scale as $N_c^1$ \cite{CG2}.
Consequently their fluctuations in a dense medium also scale as $N_c^1$.
Since the scaling of the fluctuations of conserved charges in the hadron gas
is $ N_c^0$ and it is $ N_c^1$  in the stringy fluid, one observes a clear change of the scaling $N_c^0 \rightarrow N_c^1$ just across
the chiral transition, see the lattice data in Fig.~\ref{fluc}.
 
\section{Structure of the stringy fluid}

The microscopic structure of the stringy fluid can be shortly described
in the following terms \cite{GNW}: 

$\bullet$ It is a densely packed system of the overlapping color-singlet
objects.
 
$\bullet$ There are no deconfined gluons.

$\bullet$ The chiral symmetry restoration in the confining regime happens because of the Pauli blocking of the quark levels, required for the existence of the quark condensate, by the thermal excitations of quarks and antiquarks.

$\bullet$ It is a highly collective medium with a very small mean-free
path of the color-singlet constituents.

$\bullet$ The propagating in time degrees of freedom are only color-singlets,
in which the massless quarks and antiquarks are connected by the chromoelectric string.

$\bullet$ The quark interchanges between the overlapping color-singlet 
clusters, required by the Pauli principle, make the
non-propagating in time fluctuations of the conserved charges looking as
if the quarks were quasi-free.

\section*{Acknowledgments}
The  support through the grant PAT3259224 of the Austrian Science Fund (FWF) is acknowledged.

\end{document}